# Capacitated Team Formation Problem on Social Networks

Samik Datta
Bell Labs, Bangalore
samik.datta@alcatel-lucent.com

Anirban Majumder
Bell Labs, Bangalore
anirban.majumder@alcatel-lucent.com

KVM Naidu*
Yahoo Labs, Bangalore
naidukvm@yahoo-inc.com

## ABSTRACT

In a team formation problem, one is required to find a group of users that can match the requirements of a collaborative task. Example of such collaborative tasks abound, ranging from software product development to various participatory sensing tasks in knowledge creation. Due to the nature of the task, team members are often required to work on a co-operative basis. Previous studies [1, 2] have indicated that co-operation becomes effective in presence of social connections. Therefore, effective team selection requires the team members to be *socially close* as well as a division of the task among team members so that no user is overloaded by the assignment. In this work, we investigate how such teams can be formed on a social network.

Since our team formation problems are proven to be NP-hard, we design efficient approximate algorithms for finding near optimum teams with provable guarantees. As traditional data-sets from on-line social networks (e.g. Twitter, Facebook etc) typically do not contain instances of large scale collaboration, we have crawled millions of software repositories spanning a period of four years and hundreds of thousands of developers from GitHub, a popular open-source *social coding network*. We perform large scale experiments on this data-set to evaluate the accuracy and efficiency of our algorithms. Experimental results suggest that our algorithms achieve significant improvement in finding effective teams, as compared to naive strategies and scale well with the size of the data. Finally, we provide a validation of our techniques by comparing with existing software teams in GitHub.

## 1. INTRODUCTION

Social collaboration or team formation is gaining prominence in the web. There is a growing interest in the community to understand the process of social collaboration, namely, how people use social connections to form teams in challenging situations. To understand the sociological aspects of team formation, an on-line contest was recently organized by DARPA[1]. In this contest, ten balloons were raised into the air and the participants were asked to track their locations. Since no single person could track all of them, the point of the contest was to observe how people would use their social connections to form teams and collectively report the locations.

In today's web, large scale software systems are prominent embodiments of the team formation process. In a software project, collaboration among the developers is required at multiple stages of production. Therefore, it is important to ensure that the collaboration is effective, in order to attain the highest level of productivity. Previous studies [1, 2] suggest that presence of social connections are often good indicator of effective collaboration. In fact, in our experiments (see Section 5.2), we observed a strong positive correlation between activities of a project and social nature of the corresponding team. Moreover, due to proliferation of Web 2.0 phenomena, a growing number of Open Source Software sites (e.g. GitHub [3]) are becoming 'social' by imbibing social-networking features e.g. messaging, community formation, friendship links, for their users. It remains to study how one can facilitate the team formation process in this new generation of networks by designing specialized algorithms for team selection that are mindful of the social connections of the team members. Specifically, we ask the following question in this paper: *How to build an effective team of users that meets the requirement of a project and the selected team-members are socially close to each other ?*

However, team formation is not limited to the realm of software development only. Teams are fundamental to the concept of participatory sensing [2] where a community of users contribute sensory information (through their mobile phones or other handheld devices) to constitute a body of knowledge, e.g. learning about the safety of biking tracks in a terrain [4], collecting traffic patterns in a city [5], creating map of a geographical location [6] etc. The participants of a sensing task need to collaborate among themselves in dividing the work. In this context, the social network represents their degree of co-operation and therefore, identifying the most effective community of users for a sensing task amounts to finding the 'best' team on their social network.

In summary, we believe that the problem of finding team of users on a social network (for a collaborative task), has far-reaching applications beyond the ones mentioned here and is an important problem to study. We consider the *social team formation* problem where the objective is to select a team of users from a social network and a division of the task so that the team members are socially 'close' to each other and no user is overloaded by the assignment.

---

[1]http://news.cnet.com/8301-1023_3-10410403-93.html
[2]http://en.wikipedia.org/wiki/Participatory_sensing

We would like to point out that our work is fundamentally different from the team selection problem, thoroughly studied in the Operations Research (OR) community [7, 8, 9, 10]. In a team selection problem, one is given a collection of tasks and a set of users. Given that the users have different expertise level, the goal of the team selection problem is to find an allocation of tasks to the users so as to maximize the overall expertise score. Observe that no social network is being defined between the users and the problem is being looked at merely from a resource allocation point of view. [3] Below, we summarize our contributions.

## 1.1 Our Contributions

- Our key contribution is to account for capacity constraints in the team formation problem (on a social network). For a given task, our goal is to find a team of users from the social network who are 'socially effective' in terms of collaboration (see Section 2.3) and a division of the task among team members so that no user is overloaded by the assignment i.e. no user is assigned task beyond her capacity. To the best of our knowledge, ours is the first work to formulate the team formation problem in presence of a social network and capacity constraints on users (see Section 2.2), thereby showing that it gives rise to a novel problem previously unstudied in the literature.

  This is a major departure from existing works in Operations Research [7, 8, 9, 10] that primarily model team formation as a variant of the *bi-partite matching problem* [4] or other heuristics and therefore, completely ignores the social network information. The only known works consider the social network of users are by Lappas et al [11] and Anagnostopoulos et al [12]; however, their formulations do not account for capacity constraints on the users and therefore, are fundamentally different from ours (see Section 6 for a detailed comparison).

- As our team formation problems are proven to be NP-hard, we present two efficient approximation algorithms to find nearly optimal teams with provable guarantees. Further, we consider some important extensions of our team formation problem and briefly describe how our algorithms can be adapted to handle those scenarios.

- We conduct a careful and detailed set of experiments to evaluate the performance of our algorithms. By crawling GitHub [3], a popular and the fastest growing open-source software site, we have collected large scale social networking graph and collaboration traces (hundred of thousands of users and nearly a million software repositories). Through experiments on this dataset, we have established that our algorithms achieve at much as 40% reduction in cost compared to other strategies.

  We present a validation of our algorithms by showing how our techniques can improve the collaboration cost (see Section 2.3) of existing software teams in GitHub. To the best of our knowledge, ours is the first attempt to collect and analyze large scale (in the order of hundreds of thousands of users and millions of software projects) collaboration traces from GitHub that enables us to derive some interesting insights about the nature of collaboration in this network (refer to Section 5.2).

- Further, to demonstrate generality of our techniques, we present an additional set of experiments on the DBLP dataset. Experimental evidences suggest that our algorithms achieve at least 20% reduction in cost compared to other heuristics.

## 1.2 RoadMap

Our paper is organized as follows. In Section 2, we formulate the team formation problem on social networks, that we address in this paper. Section 3 presents the algorithms for team formation and analyze their properties and performance. In Section 4, we present some key extensions of our problem and briefly describe their solution. We follow it up with detailed evaluation through experiments over real-world data-set in Section 5. We survey the related works in this area in Section 6. Finally, we conclude in Section 7.

## 2. TEAM FORMATION PROBLEM

In this section, we introduce some notations and formally define our team formation problem on a social network.

## 2.1 Notation

We represent the social network as an undirected graph $G = (V, E, w)$ with vertices in $V$ denoting the users and edges in $E$, the social connections between them. The edges are weighted by cost function $w : E \to \mathbb{R}^+$ indicating the cost of collaboration among neighboring users in the network. Therefore, for users $u$ and $v$ such that $(u, v) \in E$, if $w(u, v)$ is large, it indicates that they are unlikely to collaborate well with each other. The cost function $w$ can be estimated by analyzing previous collaboration traces of the users [5].

Each user $v \in V$ is skilled with a set of *items*, where items are, for example, programming languages, operating systems etc that are needed for a software project. We denote the set of items for user $v$ by $I_v$ and the universe of all items in the network by $\mathbb{I}$ i.e. $\mathbb{I} = \cup_{v \in V} I_v$. For user $v$, let $c_v$ denote her capacity i.e. the maximum number of items that can be assigned to $v$ without overloading the user.

Finally, a task $T$ is defined as a sub-set of items $\{i_1, i_2, \cdots i_k\} \subseteq \mathbb{I}$. A user $u$ can be assigned an item $i$ from the task only if she is skilled in it i.e. $i \in I_u$.

---
[3]On a similar note, we would like to highlight the fact that team formation or social collaboration is different from social communication, that involves messages being exchanged between a sender and a (group of) receiver in the form of e-mails, tweets etc. Social collaboration is a collective effort that requires multiple parties (i.e. users) to work together to attain a common goal.
[4]http://en.wikipedia.org/wiki/Assignment_problem

[5]Refer to Section 5.3 on details of estimating edge costs from a given collaboration trace.

## 2.2 Feasibility Constraints

A team $U \subseteq V$ is said to be feasible for a given task $T$ if it satisfies the following constraints:

- *Covering Constraints:* There is a valid assignment $\psi$ of the items in $T$ to users in $U$ such that each item $i \in T$ is assigned to some user $\psi(i) \in U$ such that $i \in I_{\psi(i)}$.

- *Packing Constraints:* For each user $u \in U$, the total number of items assigned to $u$ does not exceed her capacity i.e. $\sum_{i \in T: \psi(i) = u} 1 \leq c_u$.

## 2.3 Social Collaboration Cost

By virtue of the cost function $w(\cdot)$, it is possible to calculate the collaboration cost between any pair of users $u$ and $v$. If $u$ and $v$ are friends i.e. $(u, v) \in E$, then the cost is simply $w(u, v)$; otherwise, it is determined by the cost of the shortest path that connects $u$ and $v$. However, it is not obvious how to extend this definition of collaboration cost for a team with more than two users.

We use the following measures of collaboration cost, borrowed from the work of Lappas et al [11]

- **Diameter Cost:** Given the social network $G = (V, E, w)$, the *diameter cost* of a team $U \subseteq V$ is defined to be the maximum cost between any pair of users in $U$.

- **Steiner Cost:** Given the social network $G = (V, E, w)$, the *steiner cost* of a team $U \subseteq V$ is defined to be the minimum cost of a tree that connects all the users in $U$. Stated more formally, steiner cost for a team $U$ is the cost of a minimum cost steiner[6] tree that connects all the users in $U$.

- **Bottleneck Cost:** Given the social network $G(V, E, w)$, and a team $U \subseteq V$, the bottleneck cost is defined as the minimum cost of a tree that spans $U$, where the cost of a tree is defined as the maximum weight of any edge in the tree.

These cost models are intuitive and capture some desirable properties of a team. Observe that, while the diameter and bottleneck costs prohibit pair of users (for diameter) or neighboring users (for bottleneck cost) for whom there exists a large collaboration cost, steiner cost, on the other hand, is an aggregate measure and attempts to reduce the overall collaboration cost.

## 2.4 Problem Formulation

In a team formation problem, a user has a task in mind and is willing to collaborate with others from her social network to form a team, so that together, they accomplish the goal. Formally, it can be defined in the following way: *Given a social network $G(V, E, w)$, a user $v$ who specifies a task $T$, find a group of other users $U \subseteq V$, so that the team $U \cup \{v\}$ is feasible for the task and minimizes the social collaboration cost (measured in diameter, steiner or the bottleneck cost).*

We refer to the team formation problem with diameter collaboration cost as MINDIAMTEAM; Similarly, the variants with steiner and bottleneck collaboration costs are referred to as MINAGGRTEAM and MINMAXTEAM respectively.

In general, we assume that user $v$ also specifies a maximum hop length $h$ so that users who are more than $h$ hops away from her, are not considered by the team selection process. Typical values of $h$ are 1 (friend network), 2 (friend-of-friends network) and $\infty$ (unconstrained i.e. the entire network).

Observe that, as per our formulation, user $v$ (who creates the task) is guaranteed to be a member of the team. We believe that this scenario is a representative of social collaboration esp. in a software development environment, where the user who creates the project also contributes to it. However, this assumption is not essential to our problem, since our algorithms, presented in Section 3, can be easily adapted for the case when $v$ is not enforced to be part of the solution.

## 2.5 Hardness Result

Since the team formation problem considered in [11] is a specialization of our problem, namely, when all the user capacities are infinite, the hardness results established in their work directly carry over to our problem as well. We briefly summarize this result in the following lemma and refer to [11] for the proof. [7]

LEMMA 1. MINDIAMTEAM *and* MINAGGRTEAM *problems are NP-hard.*

## 3. ALGORITHMS

Owing to the hardness result mentioned in Lemma 1, we can not hope to compute the optimum solutions for our team formation problems MINDIAMTEAM and MINAGGRTEAM. We first present approximate algorithms to compute nearly optimum teams (with provable guarantees) under these cost models and towards the end of the section present an exact algorithm for MINMAXTEAM.

We first present an efficient procedure to identify feasible teams. Specifically, given a team of users $U$, we present techniques to determine whether $U$ is feasible w.r.t. the task at hand. We describe it in the following lemma.

LEMMA 2. MAXITEMS: *Given a task $T = \{i_1, i_2, \cdots, i_k\}$, and a team of users $U \subseteq V$, it can be computed in polynomial time, the maximum number of items that can be assigned to $U$ so that no user in $U$ is assigned items more than her capacity. (refer to Section 2.2). As a corollary, it can be decided whether $U$ forms a feasible team for the task $T$.*

PROOF. Given the task $T$ and team $U$, we form a bipartite graph $(L, R, E)$ as follows. For each item $i \in T$, we add a corresponding node $l_i$ in $L$. Similarly, for each user $u \in U$, there is a corresponding node $r_u$ in $R$. Add a directed edge $(l_i, r_u)$ of capacity 1 between $l_i$ and $r_u$ if the item $i$ can be assigned to $u$ i.e. $i \in I_u$. Augment this network by adding source and sink nodes $s$ and $t$. For each node $l_i \in L$, add a directed edge $(s, l_i)$ of capacity 1. Similarly, for each $r_u \in R$, add a directed edge $(r_u, t)$ of capacity $c_u$, i.e. the capacity of user $u$. We now seek to find the maximum flow between $s$ and $t$.

---

[6] A steiner tree $Z$ on a set of vertices $X \subseteq V$ is defined as follows - $Z$ is a tree which connects all the nodes in $X$ but might use some other vertices from $V - X$ in order to reduce the cost of connection. The cost of a steiner tree is simply the sum of the cost of its edges.

[7] However, the MINMAXTEAM problem is polynomially solvable (refer to Section 3).

**Algorithm 1** The MinDiamSol algorithm for minimizing diameter cost
---
1: **Input** Social Network $G(V, E, w)$; User $v$; Task $T = \{i_1, i_2 \cdots, i_k\}$; maximum hop-length $h$.
2: **Output** Team $U \subseteq V$; allocation of $T$ to $U$.
3: Pre-process the graph to exclude users who are more than $h$ hops away from $v$.
4: Find the smallest radius $r'$ such that MaxItems$(T_v(r')) = k$.
5: Let $Team_v$ be the team found;
6: **return** $Team_v$;
---

Observe that if there is a feasible team, then each of the edges $(s, l_i)$ will be saturated i.e. having 1 unit of flow through them. Therefore, the maximum flow is at least $k$. On the other hand, the maximum flow is bounded by the sum of the capacities of the edges $(s, l_i)$, which is $k$. Therefore, if the maximum flow of the network is $k$, we can conclude that there is a feasible team. Otherwise, it can be verified easily that the $s$-$t$ max-flow gives us the maximum number of items that can be assigned to $U$. The allocation of the items to users in $U$ can be decided by inspecting the set of edges $(l_i, r_u)$ - if the edge $(l_i, r_u)$ carries non-zero flow, then the item $i$ is assigned to user $u$.

For a graph with $n$ vertices and $m$ edges, the best known algorithm for maximum flow has $O(mn \log m)$ time complexity. □

In the following discussion, we will use OPT to denote an optimum solution[8] and |OPT|, its cost. Algorithm MinDiamSol outlines the pseudo-code of our algorithm for the MinDiamTeam problem. Our idea is to guess the value of |OPT| and decide if there is a feasible team whose cost is well approximated by our guessed value. However, we can not hope to find the optimum solution by enumerating all possible values of |OPT|, since our problem is NP-Hard. Therefore, instead of minimizing the diameter cost of the team, we rather attempt to minimize the distance between the root user (i.e. $v$) and the team members.

Assume that $r$ is our guess of the cost. We use the notation $T_v(r)$ to denote the set of users whose distance from $v$ is at most $r$ i.e. $T_v(r) = \{u|d(v, u) \leq r\}$, where $d(v, u)$ is the cost of the shortest path that connects $v$ and $u$. Observe that $v \in T_v(r)$, for all values of $r$.

Our objective is to determine the minimum value of $r$ so that the set of users $T_v(r)$ forms a feasible team for the given task.

THEOREM 3.1. *Algorithm MinDiamSol obtains a 2-factor approximation algorithm for the MinDiamTeam problem.*

PROOF. Let $\bar{r}$ be the minimum radius computed by Algorithm 3 and $r_m =$|OPT| be the cost of the optimum solution. We claim that $r_m \geq \bar{r}$. Otherwise, if $r_m < \bar{r}$, then OPT $\subseteq T_v(r_m)$ (since $v \in OPT$) implying that $T_v(r_m)$ contains a feasible team, which is a clearly a contradiction. Let $p, q$ be any two users present in the solution computed by MinDiamSol. The distance between $p$ and $q$ can be bounded in the following way,

---
[8]Observe that, as per our problem definition (Section 2.4) user v belongs to the team i.e. $v \in OPT$.

$$d(p, q) \leq d(w, p) + d(w, q)$$
$$\leq 2 \cdot \bar{r} \leq 2 \cdot r_m$$

Here the first inequality follows by applying the triangle inequality for shortest path distances. Since the diameter of the team is being defined as the maximum shortest distance between pairs of users and the inequality holds for any such pair, the proof follows. □

**Time Complexity** The feasible team MinDiamSol can be found by performing a binary search on the diameter of the graph, that requires $O(\log n_h)$ invocations of Lemma 2 where $n_h$ denotes the number of nodes that are within $h$ hops of root user $v$. Therefore, the worst-case time complexity is $O(kn_h^2(\log n_h + \log k) \log n_h)$. However, in practice, the running time of this algorithm is much less than this worst-case analysis suggests [9].

Our algorithm for MinAggrTeam problem is slightly more complicated than the diameter case. The MinAggrSol algorithm starts by transforming the social network into a simpler graph $G'$ so that finding the team in $G'$ is conceptually much easier than doing it in the original graph. Following that, the algorithm greedily picks nodes from $G'$, based on a utility measure, until a feasible team is found. These steps of MinAggrSol are shown in Algorithm 2.

**Algorithm 2** The MinAggrSol algorithm for minimizing steiner cost
---
1: **Input** Social Network $G(V, E, w)$; User $v$; Task $T = \{i_1, i_2, \cdots, i_k\}$; maximum hop length $h$.
2: **Output** Team $U \subseteq V$; allocation of $T$ to $U$.
3: Pre-process the graph to exclude users who are more than $h$ hops away from $v$.
4: $G'(V, E', \lambda) \leftarrow$ AugmentGraph(G).
5: $cover_v := \{v\}$;
6: **while** $cover_v$ **not** feasible **do**
7: $\quad \bar{x} \leftarrow \underset{x \in \{V \setminus cover_v\}}{argmax} \left\{ \frac{\text{MaxItems}(cover_v \cup \{x\}) - \text{MaxItems}(cover_v)}{\lambda(v, x)} \right\}$;
8: $\quad cover_v \leftarrow cover_v \cup \{\bar{x}\}$
9: **end while**
10: $Team_v \leftarrow$ MinSteinerTree($cover_v$, $G$); /* minimum cost steiner tree in $G$ that connects all the users in $cover_v$ */
11: **return** $Team_v$;
---

The procedure AugmentGraph (step 4 of MinAggrSol) creates a graph $G'(V, E', \lambda)$ in the following way - for each vertex $u \in V \setminus \{v\}$, it adds an edge $(v, u) \in E'$. The cost of the edge $(v, u) \in E'$ i.e. $\lambda(v, u)$ is defined as the cost of the shortest path between $u$ and $v$ in $G$. Conceptually, our construction produces a one level tree with user $v$ being the root and all other users are at level 1. A feasible team in $G'$ corresponds to a sub-tree of $G'$ with user $v$ as the root. We denote the optimum team in $G'$ by OPT$_\lambda$.

Although $G'$ is structurally much simpler than $G$, it turns out finding OPT$_\lambda$ is still NP-hard, as it can be shown by a direct reduction from the set cover problem. We therefore resort to an approximation for it. To find an approximation to OPT$_\lambda$, we start with a partial team containing user $v$ only. At each step, we greedily add a user from $G'$ to the team only if it maximizes the *utility*. The utility of a node $u \in V \setminus \{v\}$ is defined as the ratio of the marginal benefit of adding $u$ to the cover (in terms of the number of additional items it covers) to its cost $\lambda(v, u)$. We use the construction highlighted in Lemma 2 to derive the utility values. Once we find a feasible team, we construct a minimum cost steiner tree that connects all the users in the team and return it as a solution. Although computing minimum cost steiner tree is an NP-hard problem, there are well-known algorithms in the literature (c.f. [13]) that achieve factor of 2 approximation to it. The following lemma shows that the cost of the solution returned by MinAggrSol is not far from the cost of the optimum solution.

---
[9]refer to Section 5.4.3.

THEOREM 3.2. *Algorithm* MinAggrSol *obtains an $O(k \log k)$ factor approximation to the* MinAggrTeam *problem, where $k$ is the number of items required by the task.*

Our proof consists of two parts - as a first step, we find out how well OPT is approximated by $OPT_\lambda$. Following that, we prove bounds on the error we incur by approximating $OPT_\lambda$ itself. We start with the following observation.

LEMMA 3. *Cost of $OPT_\lambda$ is at most $O(k)$ times the cost of OPT.*

PROOF. Consider any optimum solution OPT=$\{v, u_1, u_2, \cdots, u_{l-1}\}$ and let $T_{OPT}$ be a minimum cost steiner tree in $G$ that connects all users in OPT. Observe that in OPT, there can be at most $k$ users who are assigned one or more items from the task i.e. $l \leq k$. In $G'$, OPT is a 1-level tree with the vertex $v$ at the root and all other vertices $\{u_1, u_2, \cdots, u_{l-1}\}$ being connected to $v$ via an edge. Observe that OPT is also a feasible solution in $G'$ whose cost is $\sum_j \lambda(v, u_j)$. Root the tree $T_{OPT}$ at node $v$. It is easy to observe that the distance of $u_j \in OPT \setminus \{v\}$ from $v$ is no more than $|OPT|$. Moreover, since $\lambda(v, u_j)$ corresponds to the shortest distance between $v$ and $u_j$ in $G$, we can write $\lambda(v, u_j) \leq |OPT|$. Therefore, the cost of OPT in $G'$ can be bounded as $(k-1) \cdot |OPT|$. Finally, observe that $OPT_\lambda$ is the minimum cost solution in $G'$ and the inequality follows. □

In the next two lemmas, we derive the factor in approximating $OPT_\lambda$.

LEMMA 4. *The function* MaxItems*(S) over set of users $S \subseteq V$ is sub-modular*[10].

PROOF. See Appendix A for the proof. □

We use a known result on the sub-modular cover problem to complete the proof of Theorem 3.2. The **sub-modular cover problem** is defined as follows: Let $E = \{1, 2, \cdots, m\}$ be a ground set and $C$ be a collection of sets defined over $E$ i.e. $C \subseteq 2^E$. Let $f$ be a sub-modular function defined as $f : 2^C \to \mathbb{Z}^+ \cup \{0\}$ such that $f(C) = m$, $f(\emptyset) = 0$ and $f$ is monotone increasing i.e. for sets $C', C'' \in 2^C$ such that $C' \subseteq C''$, it implies $f(C'') \geq f(C')$. Each set $s \in C$ has a non-negative cost $w(s)$. A sub-collection $C' \subseteq C$ is said to form a feasible cover if $f(C') = f(C)$ and the cost of the cover is defined simply as the sum of the costs of the sets in it i.e. $\sum_{s \in C'} w(s)$. In the minimum cost sub-modular cover problem one is asked to find a feasible cover of minimum cost. Wolsey et al [14, 15] obtained the following result.

[10]For a function $f : V \to \mathbb{Z}^+ \cup \{0\}$ to be sub-modular it means for all $S, T \subseteq V, f(S) + f(T) \geq f(S \cup T) + f(S \cap T)$

**Algorithm 3** The MinMaxSol algorithm for minimizing team bottleneck cost
1: **Input** Social Network $G(V, E, w)$; User $v$; Task $T = \{i_1, i_2 \cdots, i_k\}$; maximum hop-length $h$.
2: **Output** Team $U \subseteq V$; allocation of $T$ to $U$.
3: Pre-process the graph to exclude users who are more than $h$ hops away from $v$.
4: **for** $t = 0 \to maxweight(E)$ **do**
5:   Remove all edge $e \in E$ such that $w(e) > t$; Let $C$ be the component that contains $v$;
6:   If MaxItems$(C) = k$ return the team;
7: **end for**

LEMMA 5. *[14, 15] The greedy algorithm that picks sets based on maximum utility, achieves an $O(\log k)$ factor approximation to the sub-modular set cover problem.*

To put an analogy, the ground set $E$ is the set of items in our requirement $T = \{i_1, i_2, \cdots, i_k\}$ and the collection of subsets $C$ is the set of users who possesses a subset of items from $T$. Since the function MaxItems is sub-modular and satisfies all the requirements, we can conclude that through steps 8-11 of Algorithm 2) we compute an $O(\log k)$ approximate solution to $OPT_\lambda$. Combining this with Lemma 3, we obtain the result of Theorem 3.2.

**Time Complexity** Observe that in MinAggrSol, the size of $cover_v$ is at most $k$. Therefore, computing minimum cost team incurs a cost of $O(k^3 \log k + n_h k^2 + m)$ (steps 4 to 9). The steiner tree computation in step 10, can be performed in an additional $O(m + n \log n)$ time [16]. However, in practice, we find that MinAggrSol scales almost linearly with increasing size of the network (i.e. $m$ and $n$) and number of items (i.e. $k$) [11].

Algorithm MinMaxSol outlines the pseudocode of our solution for the MinMaxTeam problem. We first guess that the optimum team contains no edge with cost more than $t$ and remove edges with higher cost. Following that, we identify the component of the graph that contains the root user $v$ and determine whether it contains a feasible team. This is because any team that spans across multiple such components must use an edge of cost $> t$. Finally we search for the minimum value of $t$ such that this component (containing the root $v$) consists of a feasible team. It can be verified easily that the solution returned by MinMaxSol is optimum.

**Time Complexity** The search for minimum value of $t$ (in step 4) can be performed via binary search on the set of edges in the graph, that requires $O(\log m_h)$ invocations of Lemma 2; here $m_h$ is the number of edges in the graph after the pre-processing step (i.e. step 3 in MinMaxSol). Hence the overall time complexity is $O(kn_h^2(\log n_h + \log k) \log m_h)$.

So far we have assumed that the root user has been specified and is required to be part of the solution. However, this restriction can be lifted in the following way - we repeat the algorithms over all choices of the root node and return the team that is of minimum cost. This operation preserves the approximation factors of the algortihm but increases the time complexity by a factor of $n$. Below we present the proof for the diameter case; the result for the steiner case follows similarly.

Consider a node $v$ that belongs to the optimum solution OPT. Since we are repeating MinDiamSol for all selection of root vertices, consider the iteration when $v$ is selected as a root. In OPT, the distance of all other nodes from $v$ is at most $|OPT|$; therefore, if $r$ is picked to be equal to $|OPT|$ then the set of users $B_v(r)$ must contain a feasible team. Therefore, $r \leq |OPT|$. Now we can apply the argument in the proof of Theorem 3.1 and by observing that we are returning the minimum cost team over all selection of root vertices, the result follows.

[11]Please refer to Sections 5.4.2 and 5.4.3 for details.

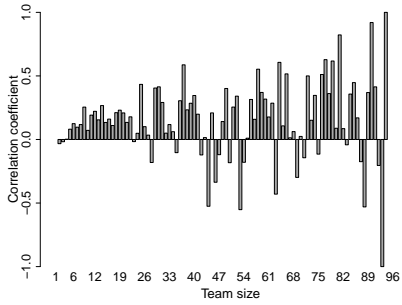 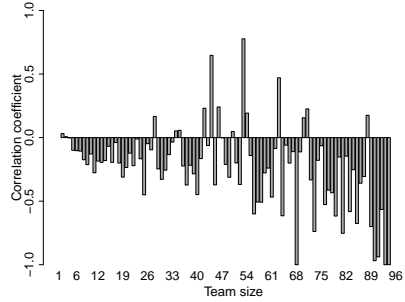 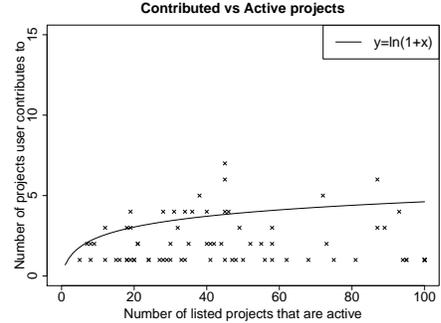

Figure 1: Correlation coefficients between number of commits of a project and number of social edges in the team for different team sizes.

Figure 2: Correlation coefficients between number of commits of a project and number of connected components in the sub-graph induced by the team-members for different team sizes.

Figure 3: Number of projects actively contributed by a user vs. number of listed projects (for 2000 users).

## 4. EXTENSIONS

In this section, we describe some key extensions of our problem and briefly outline how our algorithms can be modified to handle those cases.

**Price on users:** In a real-world situation, users are often paid for their participation in a task. In such a scenario, the objective of the team formation problem is not only to find teams that minimize the social collaboration cost but also to account for the price paid to the users. One approach could be to define the cost of a team of users $U$ in the following way: $\alpha \sum_{u \in U} \pi_u + \beta f(U)$, where $\pi_u$ is the price paid to user $u \in U$ and $f(\cdot)$ represents the social collaboration cost (see Section 2.3). Here $\alpha, \beta > 0$ are parameters to the problem. The objective is to find the team with least total cost.

In order to account for the modified cost function, we preprocess the social networking graph in the following way. For each user $u$, create a new node $u'$ and add an edge $(u, u')$ with the cost $\alpha \cdot \pi_u$ i.e. $w(u, u') = \alpha \cdot \pi_u$. In the next step, multiply each edge cost in the original graph by a factor $\beta$ i.e. the cost of the edge $(u, v)$ is set to $\beta \cdot w(u, v)$. Finally, we make the following sequence of assignments: $I_{u'} \leftarrow I_u, c_{u'} \leftarrow c_u, I_u = 0, c_u = 0$. With this pre-processing in place, we can use the algorithms MinDiamSol and MinAggrSol on the modified graph to compute the approximate solutions and the approximation factors remain unchanged.

**Multiple units of items:** We consider another generalization where a task is a multi-set of items of the form $T = \{(i_1, n_1), (i_2, n_2), \cdots, (i_k, n_k)\}$ where $\{i_1, i_2, \cdots, i_k\}$ are the set of items needed for the task and $n_j$ is the multiplicity of item $i_j$ i.e. the number of units required for item $i_j$. In this setting, a user $u$ can be assigned zero or more units of item $i$ if $i \in I_u$ and the total unit of all the items assigned to $u$ can be no more than her capacity.

We first modify Lemma 2 to account for the multiplicities of items. In the construction of the flow network, we set the capacities of the edges of the form $(l_i, r_u)$ to $n_i$, where $i$ is an item in the task and $u$ is a user. Moreover, edges of the form $(s, l_i)$ have capacities $n_i$. It can be verified that if the maximum flow has value $\sum_{i \in T} n_i$ then there is a feasible team. With this modification of the MAXITEMS procedure, we can apply algorithms MinDiamSol and MinAggrSol to obtain the approximate teams. It can be shown that while the MinDiamSol algorithm achieves a factor of 2 approximation, the MinAggrSol algorithm has an approximation ratio of $O(\log k \sum_{i \in T} n_i)$.

## 5. EXPERIMENTS

We now report a detailed set of experiments carefully designed to evaluate the performance of our algorithms. Our first step is to describe the novel data-set on which our experiments are based.

### 5.1 Dataset

Traditional data-sets gathered from popular on-line social networks (e.g. Twitter, Facebook etc.) do not contain evidence of large scale collaboration and therefore, are not suitable for our experiments. We focus our attention on a popular social coding site, GitHub [3]. GitHub hosts a large number of software projects and provide social networking features (e.g. messaging, activity streams, friendship links) for programmers working on these projects. Although there are other social coding sites (e.g. SourceForge, Google Codes etc.) we have selected GitHub for the following reasons - i) Unlike other sites, GitHub provides explicit social links through which users can connect with others, ii) GitHub APIs allowed us to extract richer quality of data and lastly, iii) it is the fastest-growing among others.

GitHub offers a novel social collaboration model in which a user can *fork* a copy of the software project owned by her friend. Contribution of the user in this forked copy can be merged with the original one, by communicating it to the owner. For our experiments, we have crawled publicly available information from GitHub consisting of 1,35,346 users and 9,05,000 software projects, during the month of August, 2011. Our data-set corresponds to 15% of the volume of GitHub and spans over a period of four years. For each project, we collected its composition in terms of program-ming languages, list of contributors, commit logs, and bug reports filed. Similarly, for each user, we collected the list of her followers and the list of the projects created and contributed by the user.

For our experiments, we crawled the following two data-sets - data-set **D1** consists of 1,35,346 users and 9,05,000 software projects and several meta-data about the users and projects (e.g. commit logs, team composition etc.). This data-set is used for estimating parameters to our algorithms (refer to Section 5.3). Our second data-set **D2**, is used as test data points. This data-set consists of new set of 470 projects collected during the period September-December 2011 (i.e. following the crawl of D1) [12].

We also perform an additional set of experiments on a DBLP data-set to demonstrate the generality of our approach. The description of this dataset and the result of our experiments are presented in Section 5.5.

## 5.2 Evidence of Social Collaboration in GitHub

We analyzed the commit logs of 10k projects over a period of last three months and observed an interesting pattern between the activity of the project (measured in terms of number of commits) and 'social' nature of the corresponding team. For this experiment, we group the teams by their size (i.e. number of contributors)[13] and for each team size, report the correlation coefficient between the social nature of the team and number of commits of the project. We use two simple measures for evaluating 'social' nature of a team - a)number of social edges present in the team and b) number of connected components in the sub-graph induced by the team members.

Figure 1 and Figure 2 show the result of this experiment. We observe that teams with more edges (hence being more social) are more active in terms of number of commits in their projects. As a result, we see positive correlation coefficient across almost all different team sizes in Figure 1. Similarly, teams with larger number of connected components (hence less social) are less active and we observe negative correlation coefficient across almost all different teams sizes (Figure 2). These results suggest that in GitHub, social nature of a team has a strong correlation with the activity of the project.

## 5.3 Preprocessing

Below, we describe in detail how data set D1 is processed to generate the input parameters to our algorithms.

**Social Network** By virtue of the follower list, we construct a social network of the users included in our crawl. Since collaboration is symmetric in nature, we treat the edges as being undirected. For each pair of users $u$ and $v$ such that $(u, v)$ is an edge, we assume their relationship strength to be proportional to the fraction of projects they have worked on together in the past. If $N_u$ and $N_v$ denote the set of projects in which $u$ and $v$ are listed as contributors, then their relationship strength is defined as $\sigma_{u,v} = \frac{|N_u \cap N_v|}{|N_u \cup N_v|}$. Finally, we use the following function to estimate the collaboration cost between $u$ and $v$: $w(u,v) = 100 \cdot (1 - \sigma_{u,v})$. [14]

---

[12] The github data can be obtained by mailing the authors.
[13] We believe that comparing number of commits of a smaller team with a larger one will be unfair. Therefore, we group teams by their size and compare among teams of same size.
[14] The constant 100 is arbitrary and is used to avoid rounding errors in our computation.

**Items** Items in data-set D1 correspond to programming languages required by the projects and there are 52 of them in our crawl. For each user, we construct her set of items as the collection of programming languages required by her projects.

**Tasks** In GitHub, tasks correspond to project requirements created by users. We use the projects in D2 and their requirements (in terms of programming languages) as tasks for our experiments. Observe that, D2 is not used for estimating other parameters since it might bias the experiment.

Further, we synthetically generate few more project requirements for our experiments. Unfortunately, a uniform random sampling of items is not guaranteed to generate realistic project requirements. Instead, we look at the co-occurrences of the items (bi-grams and n-grams in general) to generate synthetic requirements.

**Capacity** To get an understanding of capacities of users, we picked a set of ($\approx 2$k) users who are believed to be active and looked at the commit logs for their listed projects over a period of one month. Among the listed projects[15], we declare a project to be active for that user, if she made at least one commit during that one month period. The result of this experiment is plotted in Figure 3. The plot reveals an interesting trend - the number of projects contributed by a user is often significantly less compared to the number of projects listed for the user. We strongly believe that this difference is due to the limited capacities of users. Finally, we compute the capacity of user $u$ as $\ln(1 + N_u)$, where $N_u$ is the number of projects contributed by $u$, as listed by GitHub.

## 5.4 Results

### 5.4.1 Comparison with existing GitHub teams

The first set of experiments is to highlight the fact that our techniques can significantly improve the social collaboration cost of existing projects in GitHub. In order to achieve that, we worked with projects in D2 and collected their composition in terms of the list of users who contribute to them and the set of programming languages, number of units of each language needed etc. For every project, we compute the solutions returned by algorithms MinDiamSol and MinAggrSol and compare their cost against the diameter and steiner cost of the actual team. For nearly 40% of the projects, we observed a reduction of cost in the range of 10% − 35%. In Figure 5, we plot the result for the MinAggrSol. The results for MinDiamSol is similar is not repeated here.

For approximately 5% of the projects, we observed that our algorithms obtained solution which has higher cost than the actual teams. We believe that this is due our imprecise modeling of the input parameters as the test data $D2$ was kept separate and not used for estimating cost, items and capacities of the users. For rest of the projects, our algorithm reported minor improvement ($< 5\%$) over the existing teams.

### 5.4.2 Scalability

For the purpose of this experiment, we generate subgraphs of the github network of varying sizes (between 10k and 135k). The subgraphs are created by randomly selecting a node and performing breadth-first search around it. We create synthetic requirements for a randomly selected root user, through the steps described in Section 5.3, and measure the time taken by our algorithms [16]. Finally, we report the average over 10 such measurements. In Figure 7, we plot the result of this experiment. It can be seen that, Algorithms MinMaxSol and MinDiamSol are highly efficient, taking approximately 30 secs on the 135k network. Algorithm MinAggrSol on the other hand scales almost linearly with increasing network size (nearly 6 min on the 135k subgraph).

---

[15] All projects listed by GitHub are active i.e. they are being contributed to by some users in the network.

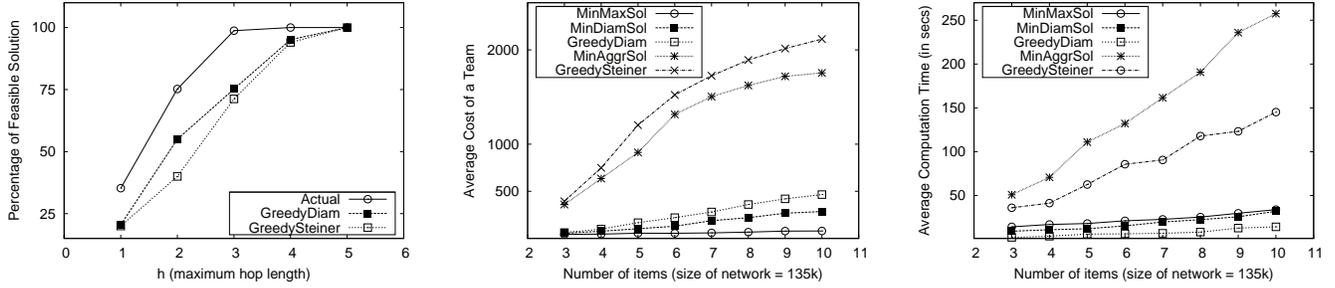

Figure 4: Experimental results for the GitHub dataset.

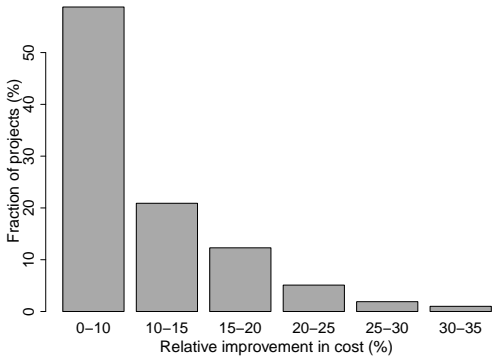

Figure 5: Improvement in collaboration cost as computed by MinAggrSol algorithm over existing teams in GitHub.

| #sketches | error | size | computation time |
|---|---|---|---|
| 1 | 51% | 10MB | 5 hours |
| 3 | 18.7% | 32MB | 18 hours |
| 5 | 9.3% | 55MB | 28 hours |

Table 1: Characteristics of the distance sketch.

### 5.4.3 Comparison with Baseline

**Baseline Algorithms** For baseline comparison, we implement an adapted version of the algorithms RarestFirst and EnhancedSteiner, from the work of Lappas et al [11]. For a given task, these algorithms find a team of experts from a social network. However, they are not designed to handle capacity constraints on the users. We modify these algorithms as follows - first, we run the algorithm and inspect its solution. If the solution is feasible (c.f. Section 2.2), we declare it to be valid. Otherwise, for any user $u$ whose ca-pacity constraint is not met, we perform a reallocation of items. Among the items that are assigned to $u$, we order them based on their *rarity* - the minimum increase in cost to cover the item by other users who are not included in the current solution, and greedily assign the top $b_u$ rarest items to $u$. After we have performed reallocation for all violating users, we contract the set of users into a single node of the graph and repeat the step, until a feasible solution is attained. We name these algorithms GreedyDiam and GreedySteiner respectively.

---

[16] For this experiment, we set $h$ to $\infty$; i.e. the algorithms are run on the entire subgraph.

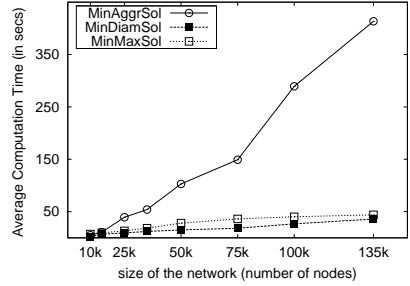

Figure 7: Computation time of different algorithms with varying network size.

**Optimization** Through some initial set of experiments, we observed that MinDiamSol is quite efficient in terms of processing time with the average being ≈ 25 seconds. On the other hand, a straightforward implementation of the MinAggrSol algorithm does not directly scale with the data. A sample run of MinAggrSol on takes close to 25 minutes to complete. Profiling of MinAggrSol revealed that a significant amount of time is spent on computing shortest path distances between nodes of the graph (step 4 and 10 in MinAggrSol). We therefore implement a distance sketch that provides quick estimation of shortest path queries.

We have borrowed the idea of pre-computing approximate shortest path distances from the work of Das Sharma et al [17]. The approach is to maintain a sketch structure that consists of a small number of *landmark* nodes and precompute the shortest path distances through them. Our experiments suggest that by storing only 5 sketches, one can reduce the average relative error below 10% (see Table 1). The total memory footprint of the sketch for our data-set is 55MB. With the optimization in place, a sample run of of MinAggrSol takes only 380 seconds in average.

**Results** Our current set of experiments is designed to evaluate the performance of MinDiamSol and MinAggrSol with respect to the baseline heuristics. We first determine the minimum value of $h$, the maximum hop length, for which our input tasks have feasible solution. In Figure 4, we plot the result of this experiment (leftmost plot). As it can be seen, for a value of $h = 3$, our tasks (both real and synthetic) have feasible solutions. However, it turns out that both GreedyDiam and GreedySteiner fail to find feasible solutions for all of the tasks unless we increase $h$ to 5. We therefore set $h$ to 5 for the rest of the experiments.

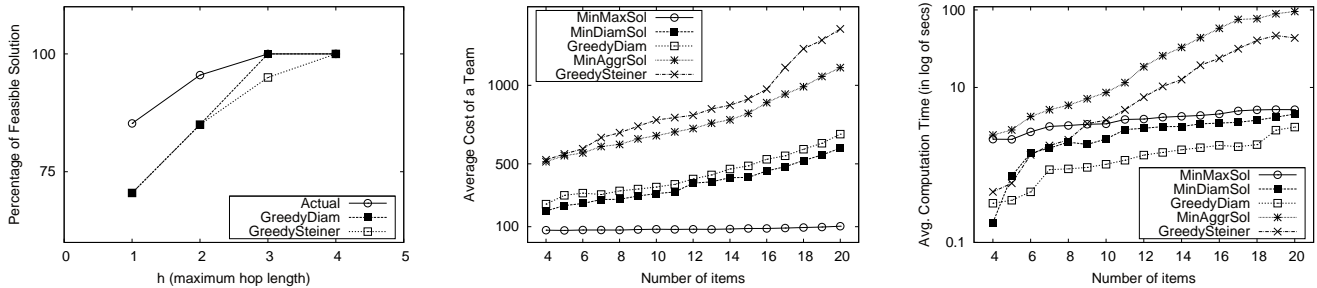

Figure 6: Experimental results for the DBLP dataset.

Figure 4 summarizes the result of our experiment with real tasks. The experiments reveal that MinDiamSol achieves at least 25% reduction in cost as compared to GreedyDiam. The improvement is much more pronounced in the steiner cost version where our algorithm significantly outperforms the GreedySteiner heuristic by at least 40%. Our careful inspection of the steps of GreedySteiner reveals that by ignoring capacities of users in its decision making process, it actually ends up including a large number of steiner nodes in the solution and thereby, increasing the cost. The plot on running time (Figure ??) indicates that our algorithms scale linearly with the number of items in a task.

In Figure 8, we plot the result of our experiment with synthetic tasks. For the synthetic data-set, our algorithms achieve 30% reduction in cost. The slight reduction in performance factor can be attributed to the fact that a large fraction of synthetic tasks contain only 3 items. As Figure 8 shows, for tasks with only 3 skills, both the baseline and our algorithms have comparable performance and the improvement is progressively better for larger tasks.

For a given task, it might happen that there is no feasible team of users that defines a connected sub-graph. Even if such a team exists, it could also be possible that our algorithms fail to find it. Since disconnected teams are bad for collaboration, it is imperative to evaluate how often our algorithms report them. In Figure 8, we plot the fraction of disconnected teams returned by our algorithms and the baseline. It is to be observed that 18% of teams reported by MinDiamSol are disconnected. We strongly believe that these are the tasks for which there can be no connected team in the input graph. Both MinDiamSol and GreedyDiam find approximately same fraction of disconnected teams. GreedySteiner, on the other hand, returns disconnected teams substantially more often than MinAggrSol.

## 5.5 Experiments on DBLP data

Our second data-set is procured from DBLP [18] and consists of a snapshot of computer science publications taken on December 5th, 2011. Each publication is written by a set of authors. We follow the same pre-processing steps as outlined in Lappas et al [11] to generate input parameters for our algorithms. In addition, we determine capacity of an author $u$ as the average number of publications co-authored by $u$ in a year.

Unlike GitHub, DBLP dataset doesn't have an explicit notion of social graph. We generate the social network based on co-authorship: two authors are connected by an edge if they have co-authored at least two papers together. The collaboration cost of an edge $(u, v)$ is estimated in the following way: If $S_u$ ($S_v$) denotes the set of research articles co-authored by $u$ ($v$ resp.) then the cost of edge $(u, v)$ is estimated as $100 \cdot (1 - \frac{|S_u \cap S_v|}{|S_u \cup S_v|})$. Once again, the constant 100 is arbitrary and is used to avoid any rounding error in computation. Following the pre-processing steps, the DBLP dataset contains 7332 authors, 19248 articles and 2763 distinct items.

### 5.5.1 Results

The result of our experiment is summarized in Figure 6. Since the DBLP network is slightly denser than GitHub, the baseline algorithms can find feasible teams within 4 hops of the root user. Therefore, for the purpose of this experiment, $h$ is set to 4. We observe that MinDiamSol and MinAggrSol achieve $\approx 20\%$ improvement over the baseline heuristics. As mentioned earlier, our synthetic construction of the DBLP network is slightly denser than GitHub and contains users with large number of publications (hence large capacity). This fact has led to slight reduction of the gains reported by our algorithms as compared to GitHub experiments.

To conclude, through a fairly extensive set of experiments on real-world datasets, we have evaluated the performance of our team formation algorithms. Experimental results suggest that our algorithms outperform other naive strategies by a significant margin (20-40%) and scale well with hundreds of thousands of users.

## 6. RELATED WORK

Team formation is a well-studied problem in the Operations Research (OR) community [9, 10, 7, 8]. In their setting, the goal is to find an allocation of tasks to different individuals so as to satisfy the project requirements. However, this body of work ignores the social relationship among people while deciding the best team and unfortunately, their techniques can not be applied to solve our problem.

Lappas et.al [11] considered the problem of finding team of experts on a social network. Given a requirement of skills, their goal is to find a team that minimizes the overall communication cost. Although they have a similar objective in mind, our problem is fundamentally very different from them. In [11], the authors don't deal with capacities on users. In effect, their problem can be thought of as *uncapacitated* version of our problem. As per our understanding, the notion of capacity changes the problem structure completely and their solutions can no longer be adapted to handle our requirement. Moreover, through detailed experiments (see Section 5), we have established that a natural extension of their algorithms perform very poorly compared to ours.

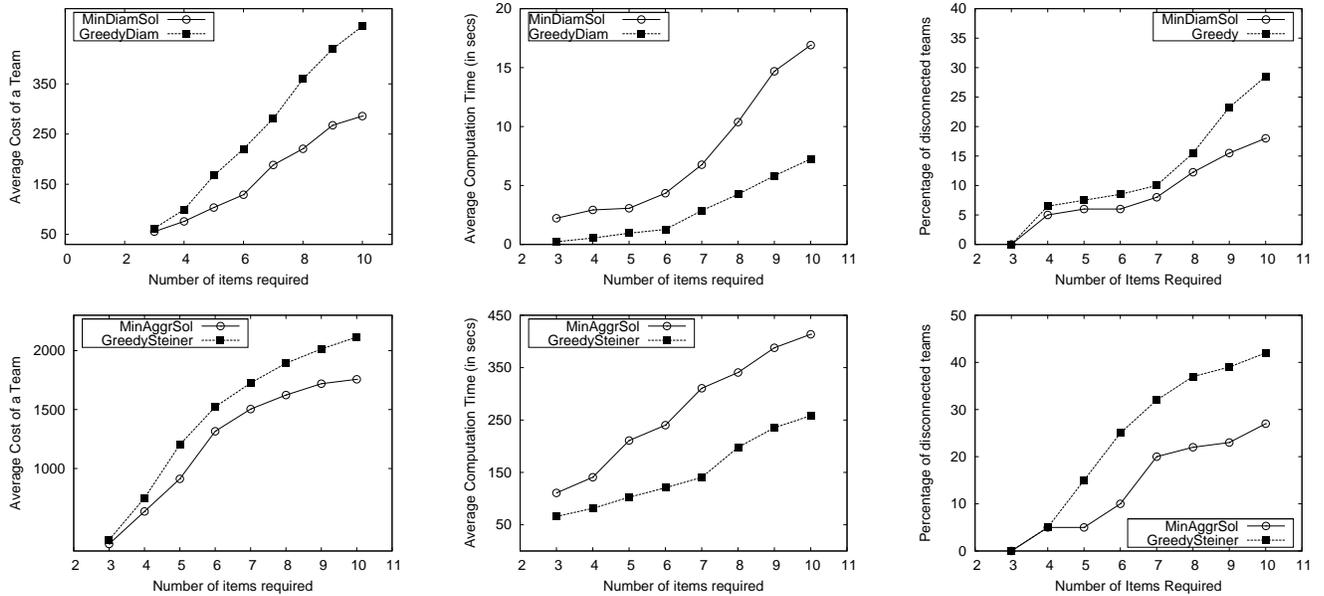

Figure 8: Experimental results on GitHub dataset with synthetic tasks.

In [19], the authors studied a version of the team formation problem where the goal was to balance the load (*i.e.* the number of projects one is assigned to) of individuals. However, this version of the problem overlooks the social collaboration cost altogether. In a more recent study [12], the authors present algorithms which, in addition to load balance, also find teams with bounded social collaboration cost. Their formulation is, however, different from ours - in addition to finding teams with bounded collaboration cost, they attempt to minimize the load on team members, whereas, we take the user capacities as constraint and attempt to minimize the social collaboration cost.

The specific techniques required to estimate the strength of relationship among individuals is an active and important area of research that has been pursued in other research works [20, 21]. However, we believe that this direction of research is orthogonal and beyond the scope and goal of this paper.

## 7. CONCLUSION

In this work, we presented a novel team-formation problem on social networks that attempts to find teams that are socially close as well as division of the task so that no user is overloaded by the assignment. We presented algorithms for this problem under different models of social collaboration and analyzed their performance thoroughly. Our experiments with real-life data-sets show that these techniques outperform the naive strategies by a significant margin.

## 8. REFERENCES

[1] D. McDonald, "Recommending collaboration with social networks: A comparative evaluation," in *CHI'03*, 2003.
[2] T. Wolf, A. Schroter, D. Damian, L. Panjer, and T. Nguyen, "Mining task-based social networks to explore collaboration in software teams," *Software, IEEE*, 2009.
[3] GitHub, "http://github.com."
[4] The OpenCycleMap project, "http://www.opencyclemap.org/."
[5] A. Bayen *et al.*, "Mobile Millennium Final Report," UC Berkeley, Tech. Rep. UCB-ITS-CWP-2011-6, 2011.
[6] The OpenStreetMap project, "http://www.openstreetmap.org/."
[7] A. Zzkarian *et al.*, "Forming teams: an analytical approach," *IIE Transactions*, 1999.
[8] H. Abdelsalam, "Multi-objective team forming optimization for integrated product development projects," in *FOCI Volume 3*, 2009.
[9] A. Baykasoglu *et al.*, "Project team selection using fuzzy optimization approach," *Cybern. Syst.*, 2007.
[10] H. Wi *et al.*, "A team formation model based on knowledge and collaboration," *Expert Syst. Appl.*, 2009.
[11] T. Lappas *et al.*, "Finding a team of experts in social networks," in *KDD '09*, 2009.
[12] A. Gionis *et al.*, "Online team formation in social networks," in *WWW '12*, 2012.
[13] V. V. Vazirani, *Approximation Algorithms*. Springer, 2004.
[14] L. Wolsey, "An analysis of the greedy algorithm for the submodular set covering problem," *Combinatorica*, vol. 2, pp. 385–393, 1982.
[15] G. Nemhauser and L. Wolsey, "Maximizing submodular set functions: Formulations and analysis of algorithms," in *Annals of Discrete Mathematics*

*(11) - Studies on Graphs and Discrete Programming.*
North-Holland, 1981.
[16] K. Melhorn, "A faster approximation algorithm for steiner tree problem in graphs," *Inf Process. Lett.*, vol. 27, pp. 125–128, 1988.
[17] A. Das Sarma *et al.*, "A sketch-based distance oracle for web-scale graphs," in *WSDM '10*, 2010.
[18] DBLP, "http://dblp.uni-trier.de."
[19] A. Anagnostopoulos *et al.*, "Power in unity: forming teams in large-scale community systems," in *CIKM '10*, 2010.
[20] R. Xiang *et al.*, "Modeling relationship strength in online social networks," in *WWW '10*, 2010.
[21] J. Leskovec *et al.*, "Predicting positive and negative links in online social networks," in *WWW '10*, 2010.

# APPENDIX

## A. PROOF OF LEMMA 4

PROOF. Observe that MAXITEMS is a monotone function i.e. for $S \subseteq T$, MAXITEMS$(S) \leq$ MAXITEMS$(T)$. In order to prove sub-modularity, we need to establish the following result: for two arbitrary subsets $A$ and $B$, MAXITEMS$(A)$ + MAXITEMS$(B) \geq$ MAXITEMS$(A \cup B)$ + MAXITEMS$(A \cap B)$. As a first step, we prove it for sub-sets $A$ and $B$ such that $A \cap B = \emptyset$.

For a given task $T$, consider the allocation of items by MAXITEMS on the set of users $A \cup B$. Let $n_A$ and $n_B$ denote the number of items assigned to sets $A$ and $B$ respectively. Since MAXITEMS(A) assigns the maximum possible items to $A$ we can claim $n_A \leq$ MAXITEMS$(A)$ and similarly, $n_B \leq$ MAXITEMS$(B)$. The proof follows by observing that MAXITEMS$(A \cap B) = 0$ since $A$ and $B$ are disjoint.

For the general case, consider an allocation of the given task $T$ to the set of users $A \cup B$ by MAXITEMS$(A \cup B)$. Let $n_A$, $n_B$ and $n_{A \cap B}$ denote the number of items assigned to sets $A$, $B$ and $A \cap B$ respectively. Therefore, $n_A + n_B - n_{A \cap B} =$ MAXITEMS$(A \cup B)$. Applying a similar reasoning (as in the last paragraph) we get, MAXITEMS$(A)$ + MAXITEMS$(B) \geq$ MAXITEMS$(A \cup B) + n_{A \cap B}$. Now the only thing needs to be proved is whether there exists some solution of MAXITEMS$(A \cup B)$ such that $n_{A \cap B} =$ MAXITEMS$(A \cap B)$. This can be achieved by constructing a minimum cost flow network in the following way: In the construction of Lemma 2, consider all edges of the form $(l_i, r_u)$ where $i$ is an item and $u \in A \cup B$. The edge is assigned a cost zero only if $u \in A \cap B$. For all other users, the cost is set to 1. For other set of edges, the cost is set to zero. Finally, we compute a minimum cost flow of value MAXITEMS$(A \cup B)$. It is easy to verify that this construction produces a re-allocation of MAXITEMS$(A \cup B)$ so that $n_{A \cap B} =$ MAXITEMS$(A \cup B)$ and this completes our proof. □